\documentclass[a4paper,11pt]{article}
\usepackage{graphicx}
\usepackage{amsmath}
\usepackage[varg]{txfonts}
\usepackage{indentfirst}
\usepackage{url}
\usepackage{authblk}
\begin{document}

\title{Unobservability of the topological charge in nonabelian gauge theory: Ward-Takahashi identity and phenomenological aspects}
\author[1,2]{Nodoka~Yamanaka\footnote{nyamanaka@kmi.nagoya-u.ac.jp}}
\affil[1]{Kobayashi-Maskawa Institute for the Origin of Particles and the Universe, Nagoya University, Furocho, Chikusa, Aichi 464-8602, Japan}
\affil[2]{Nishina Center for Accelerator-Based Science, RIKEN, Wako 351-0198, Japan}

\date{\today}
\maketitle

\abstract{
We argue that the topological charge of nonabelian gauge theory is unphysical.
To show this statement, we use the Adler-Bardeen theorem and the Becchi-Rouet-Stora-Tyutin symmetry which are warranted by the perturbative finiteness of the chiral anomaly, thus being free of Gribov ambiguity.
In addition to the original argument using the unphysical gauge field component collinear to the spatial derivative of the gauge function, we show the unobservability of the topological charge using the Ward-Takahashi identity.
We then present the consequences of this finding and show the consistency with many physical pictures and ideas that have been developed around the topology of nonabelian gauge theory.
The most important ones are the resolution of the Strong CP problem, the unobservability of topological instantons, the physical relevance of the axial $U(1)$ symmetry, the independence of the vacuum energy on the vacuum angle, and the impossibility to realize the sphaleron induced baryogenesis and chiral magnetic effects.
The axial $U(1)$ symmetry and the unphysical $\theta$-term imply that the physical complex phase of the Cabibbo-Kobayashi-Maskawa matrix is the sole source of CP violation in the standard model.
The unphysical sphaleron also means that the lepton number is phenomenologically free from the baryon number, and their violations may be modeled separately.
We also comment on the consistency with the results of lattice calculations.
}

\tableofcontents

\section{Introduction}

Adler, Bell and Jackiw found an anomalous nonconservation of the axial current in the triangle diagram generated in chiral $U(1)$ gauge theories \cite{Adler:1969gk,Bell:1969ts,Adler:2004qt}.
It was then successfully incorporated in the path integral formalism by Fujikawa \cite{Fujikawa:1979ay,Fujikawa:1980eg}.
Similarly, we can derive the topological charge density of nonabelian gauge theory 
\begin{equation}
\frac{\alpha_s}{8\pi}
F_{\mu \nu, a}\tilde F^{\mu \nu}_a
\equiv F \tilde F
,
\label{eq:topological_charge_density}
\end{equation}
where $ F^{\mu \nu}_a$ and $\tilde F^{\mu \nu}_a \equiv \frac{1}{2}\epsilon^{\mu \nu \rho \sigma}F_{\rho \sigma ,a}$ are the gauge field strength and its dual, respectively, and $\alpha_s\equiv \frac{g_s^2}{4 \pi}$ is their coupling constant.
This quantity is constructed by forming the triangle diagram with a global $U(1)$ current, for which we take the divergence, and two other nonabelian local ones [e.g. $SU(N_c)$], with one or all of them chiral.
This type of quantum anomaly involves a global current, so it is physically observable, but it has the important property that it becomes unphysical when the global symmetry is promoted to a local gauge symmetry so that we need a compensating charged field to keep it physical, and this feature holds at any arbitrary scale (anomaly matching of 't Hooft \cite{thooftanomalymatchingconodition}).
Equation (\ref{eq:topological_charge_density}) is a total divergence, but its space-time integral may be nonzero due to the topologically nontrivial entanglement of the nonabelian gauge degrees of freedom with the space.
Formally, it is called ``abelian anomaly'', because the current which does not conserve due to the presence of Eq. (\ref{eq:topological_charge_density}) is axial $U(1)_A$ (it may also be vector $U(1)_V$, if one of the nonabelian currents is axial).
The integral of Eq. (\ref{eq:topological_charge_density}), the topological charge, is indeed
\begin{equation}
\int d^4 x \,
F \tilde F
=
\frac{i g_s \alpha_s}{24\pi}
\int d^3 x \,
f_{abc}
\epsilon_{ijk}
A_{i,a}(x)A_{j,b}(x)A_{k,c}(x)
\Bigg|^{t=+\infty}_{t=-\infty}
=
\Delta n
,
\label{eq:topological_charge}
\end{equation}
where $\Delta n$ is an integer given by the shift of the winding number between the nonabelian gauge fields $A^\mu_a(x)$ at $t\to -\infty$ and $t\to +\infty$.
The topological charge is then the operator which extracts the topological information from the ``winding'' background gauge field.

After the discovery of the instanton in Yang-Mills theory \cite{Belavin:1975fg,Schafer:1996wv}, 't Hooft suggested that it is the cause of the explicit $U(1)_A$ violation \cite{tHooft:1976rip,tHooft:1976snw}.
The instanton was so far not been discovered in experiments \cite{Khoze:2019jta,Khoze:2020tpp,Khoze:2020paj}, but its existence is suggested from lattice simulations \cite{Chu:1994vi,Faccioli:2003qz,Athenodorou:2018jwu}.
Witten and several other authors discussed the consistency of the $U(1)_A$ problem \cite{Weinberg:1975ui} in the large $N_c$ expansion, and contested the instanton as the solution to the $U(1)_A$ problem (due to the heavy would-be Nambu-Goldstone (NG) boson for the $U(1)_A$ symmetry) \cite{Witten:1979vv,Veneziano:1979ec,Veneziano:1980xs}.
Another approach to resolve the $U(1)_A$ problem, consisting of showing the unphysicalness of the NG boson of the $U(1)_A$ symmetry, was also taken by several other authors.
Kogut and Susskind proposed the dipole mechanism, which is relevant in the 1+1 dimension QED (Schwinger model), to explain the absence of the light ninth NG boson \cite{Kogut:1973ab,Kogut:1974kt}.
Kugo improved this discussion by promoting the dipole mechanism to the quartet mechanism, where the topological current is unphysical by being a component of the Becchi-Rouet-Stora-Tyutin (BRST) quartet \cite{Kugo:1978nc,Kugo:1979gm,Hata:1980hn}.

The existence of the instanton leads to the nontrivial topological structure of the vacuum.
The two-point correlator of an operator $D$ carrying a topological charge $n'$ when the spatial coordinates are infinitely separated behaves like
\begin{equation}
\langle n | D (x) D (y) |n \rangle
\hspace{1.em} 
\to
\hspace{-1.5em} 
_{{}_{\tiny |x-y|\to \infty }}
\hspace{+1.em}
\langle n| D (x) | n+n' \rangle \langle n+n' | D (y) |n \rangle
\ne 0
,
\label{eq:topologicalclusterdecomposition}
\end{equation}
where the vacuum $|n \rangle $ belongs to the $n$-th topological sector.
As we may see, the cluster decomposition does not work, and this fact lead to claim that vacua with fixed topology are not physical. 
To resolve this problem, the coherent superposition of all vacua with different topology $|n\rangle $, the so-called $\theta$-vacuum, 
\begin{equation}
|\theta \rangle
=
\sum_{n = -\infty}^{+\infty}
e^{i n \theta }
| n \rangle
,
\end{equation}
was introduced \cite{Jackiw:1976pf,Callan:1976je}.
The effect of topology changing operators then just reduces to a complex phase rotation.
This vacuum satisfies the cluster decomposition even when fermions are coupled to the gauge theory, and it may be translated to a theory with a trivial vacuum $|0\rangle$ and with a Lagrangian containing the $\theta$-term
\begin{equation}
{\cal L}_\theta
=
\theta
F\tilde F
.
\label{eq:theta-term}
\end{equation}
This nontrivial vacuum is however parity and CP violating except for $\theta = 0, \pi$, and has observable effects such as the electric dipole moment (EDM) of the neutron \cite{Crewther:1979pi,Pich:1991fq,Mereghetti:2010kp,deVries:2015una,Shindler:2015aqa,Dragos:2019oxn}.
The latest experimental result of the neutron EDM, $|d_n |< 1.8 \times 10^{-26} e$ cm \cite{Abel:2020gbr}, can be converted to the constraint $|\theta | < 10^{-10}$, under the assumption that $\theta$ is the sole source of CP violation.
There are also other strong constraints from the experimental data of the EDM of diamagnetic atoms \cite{Graner:2016ses,Yamanaka:2017mef,Sachdeva:2019rkt} as well as from the decay of $\eta$ and $\eta'$ mesons \cite{Crewther:1979pi,Kawarabayashi:1980dp,Kawarabayashi:1980uh,Aaij:2016jaa,Babusci:2020jwb,Bass:2018xmz,Gan:2020aco}.
This extreme fine-tuning of the vacuum angle is nowadays called the ``Strong CP problem'' \cite{Kim:2008hd}, and several interesting resolutions have been conceived.
The first possibility is to have at least one massless quark, which will allow us to perform $U(1)_A$ transformation so as to unphysicalize the $\theta$-term (\ref{eq:theta-term}) through the anomalous chiral Ward-Takahashi identity (WTI), but this case is hardly ruled out by recent lattice calculations \cite{Alexandrou:2020bkd}.
The second and the most popular scenario is the axion mechanism \cite{Peccei:1977hh}, which consists of adding a new scalar field coupled to the topological charge of QCD.
The vacuum expectation value of this scalar field exactly cancels the effect of the $\theta$-term, but this instead gives rise to a very light observable mode, the axion \cite{Weinberg:1977ma,Wilczek:1977pj}.
The search for axions is currently one of the hottest topic of particle physics, but no such particles have been found so far \cite{DiLuzio:2020wdo,XENONCollaboration:2022kmb}.
There are also other interesting resolutions such as the generation of finite $\theta$ angle through spontaneous CP breaking \cite{Nelson:1983zb,Barr:1984qx,Barr:1984fh}, or the decoherence of the topological sectors due to the horizon of the expanding Universe \cite{Torrieri:2020nin}.
Other than these attempts to find physical solutions, there were also conceptual discussions on the gauge symmetry and the discontinuity of the topological sectors \cite{Healey:2010pss,Dougherty:2020zdl,Gomes:2020miz}.

This paper is devoted to explain the resolution of the Strong CP problem by showing that the topological charge of nonabelian gauge theory is actually unphysical, and to summarize important changes in particle physics.
In the next section, we present the synopsis of the resolution already presented in Ref. \cite{Yamanaka:2022vdt}.
We also add a short comment on the Schwinger model.
In Sec. \ref{sec:BRST}, we give a more robust derivation of the unobservability of the topological charge based on the WTI and the BRST symmetry.
We then inspect in Sec. \ref{sec:fermion} the chiral WTI in the presence of fermions.
In Sections \ref{sec:QCDsym} and \ref{sec:strocchi}, we explain the consistency of the symmetry breaking pattern of massless QCD with the spontaneous breaking of $U(1)_A$ and that of our mechanism with the cluster decomposition theorem given by Strocchi, respectively.
In the remaining part of this paper we will focus on the phenomenological aspects, namely, the Dirac zero-modes and multi-fermion vertices (Sec. \ref{sec:zero-modes}), the topological susceptibility and $\eta'$ mesons (Sec. \ref{sec:etaprime}), instanton models (Sec. \ref{sec:instanton}), the phenomenology of axions (Sec. \ref{sec:axion}), phenomenological constraints on physics beyond the standard model (SM) from EDM experiments (Sec. \ref{sec:EDM}), and the consistency with lattice calculations (Sec. \ref{sec:lattice}).
The final Section summarizes all topics explored in this paper.

\section{General inspection of the topological charge\label{sec:topologicalinspection}}

In this Section, we review the discussion of Ref. \cite{Yamanaka:2022vdt} which showed that the topological charge is unphysical.
In looking at Eq. (\ref{eq:topological_charge}), one remarks that we need to have three degrees of freedom for the polarization of the gauge field, since the time dependence is frozen at $t\to \pm \infty$.
The topological charge only involves spatial components, so it also covers the direction of the gradient of the gauge function $\chi_a (x)$ which appears in the local gauge transformation
\begin{equation}
A_a^\mu (x) \to A_a^\mu (x) + \partial^\mu \chi_a (x) + O(g_s)
.
\label{eq:localgaugetransform}
\end{equation}
We then anticipate that it is an unphysical quantity.
The gauge component proportional to the gradient of $\chi_a (x)$ is unphysical in perturbative quantum field theory according to the BRST symmetry \cite{Becchi:1975nq,Tyutin:1975qk,Gupta:1949rh,Bleuler:1950cy,Nakanishi:1972pt,Kugo:1977zq,Kugo:1979gm}, which replaces the local gauge symmetry after gauge fixing.
The $O(g_s)$ terms are irrelevant in perturbation theory, and this fact is quite important in our discussion.
The object of this Section is to show that the topological charge is irrelevant in observables and that the topological information is not physical.

We now concretely demonstrate that the topological charge always involves a component proportional to the gradient of the gauge function.
In the first equality of Eq. (\ref{eq:topological_charge}), there is a triple product (contraction with $\epsilon_{ijk}$) of the spatial components of the gauge field.
It is well known that $\epsilon_{ijk}$ must be contracted with three linearly independent vectors, so the triple product is proportional to all spatial components of the gauge field, and this must also cover the direction of the gradient $\vec \nabla \chi_a (\vec x)$.
In this discussion, the time coordinate is irrelevant because the temporal integral has already been performed in Eq. (\ref{eq:topological_charge}), and we may restrict the discussion to the three-dimensional space.
Since the triple product covers all three-dimensional directions, it is evident that the topological charge is proportional to $\vec \nabla \chi_a (\vec x)$, which is unphysical according to the BRST symmetry.

We worked so far in the perturbation theory where the BRST symmetry is certified.
It is however known that the latter does not hold when the Gribov ambiguity \cite{Gribov:1977wm,Singer:1978dk,Zwanziger:1989mf,Maggiore:1993wq,Dudal:2008sp,Vandersickel:2012tz} is taken into account.
Nonabelian gauge theories are known to not be unique when Lorentz covariant gauge fixing is performed, and there are still remaining redundant configurations satisfying the same gauge fixing condition while belonging to the same gauge orbit, also called Gribov copies.
Gauge transformations change the amplitude of gauge fields, and Gribov copies always appear when it is sufficiently increased (here "amplitude" is not the scattering amplitude, but that of the gauge field).
We then have to restrict the path integral to a region where the field amplitude does not contain any copies.
Such a perfect restriction (called fundamental modular region) has not been formulated yet, but the first Gribov region, the restriction based on the positivity of the Faddeev-Popov determinant, which is a product of all eigenvalues of the Faddeev-Popov operator
\begin{equation}
-\partial_\mu D_\mu 
,
\label{eq:FP_operator}
\end{equation}
in the Euclidean space-time (imaginary time), 
where $D^\mu = \partial^\mu +g_s f_{abc} A_b^\mu$ is the covariant derivative, has actively been studied until today since the first work of Gribov \cite{Gribov:1977wm,Singer:1978dk,Zwanziger:1989mf,Dudal:2008sp,Vandersickel:2012tz}.
It is actually known that the BRST symmetry breaks down under this restriction \cite{Maggiore:1993wq}, and also that gauge bosons dynamically get mass \cite{Dudal:2008sp,Vandersickel:2012tz}.
The massive behavior was also seen in lattice calculations of Landau gauge QCD  \cite{Iritani:2009mp,Dudal:2018cli,Falcao:2020vyr}.

The breakdown of BRST symmetry seems to falsify our assumption, but the case of the topological charge is however an exception.
It is actually possible to show that the BRST symmetry holds for the gauge field operators of the topological charge, and to consequently claim that the latter is unobservable.
According to Adler-Bardeen theorem, the chiral anomaly, and the topological charge which is its integral, are not renormalized by quantum corrections (up to the field strength renormalization), and their effects are strictly perturbative, only given by the one-loop level contribution  \cite{Adler:1969er,Adler:1972zd,Higashijima:1981km,Costa:1977pd,Fujikawa:1981rw,Lucchesi:1986sp,Anselmi:2014kja,Anselmi:2015sqa,Mastropietro:2020bhz}.
The Adler-Bardeen theorem is just a synonym of 't Hooft anomaly matching \cite{thooftanomalymatchingconodition} and it holds at any scale even if nonperturbative effects are present. 
In perturbation theory, the amplitude of a field is infinitesimal.
This may be understood by writing Glauber's formulation of coherent fields \cite{Glauber:1963tx}
\begin{equation}
| \epsilon \rangle
=
e^{-\frac{1}{2}|\epsilon|^2}
\sum_n
\frac{(\epsilon a^\dagger)^n}{n!}
| 0 \rangle
=
(1+ \epsilon a^\dagger)| 0 \rangle
,
\end{equation}
where only the first power of field operator $a^\dagger$ with the infinitesimal amplitude $\epsilon$ contributes to the physics (the first term with $1$ is not physical since it does not create any moving fields).
With such gauge fields, the eigenvalues of the Faddeev-Popov operator (\ref{eq:FP_operator}) cannot be negative, since the gauge interaction term of the covariant derivative becomes infinitesimally small, and we do not have Gribov copies (same situation as the quantum electrodynamics which does not have gauge self-interaction).
In other word, infinitesimal fields always lie in the fundamental modular region, so the BRST symmetry must be exact there.
The contributions of the chiral anomaly and the topological charge to the path integral are therefore free of Gribov ambiguity thanks to the perturbative finiteness.
This justify the assumptions used in our statement, that the topological charge is unphysical because it always involves a gauge field component proportional to the gradient of the gauge function, and we may therefore claim that the topological charge is unobservable.

The unobservability of the topological charge immediately leads to the resolution of the Strong CP problem since Eq. (\ref{eq:theta-term}) is also unphysical.
We emphasize that the $\theta$-term is strictly unphysical, even at the classically CP conserving point $\theta = \pi$.
At $\theta = \pi$, it is known that the CP symmetry breaks spontaneously (so-called Dashen's phenomenon) \cite{Dashen:1970et,Gaiotto:2017yup}, but this will have no physical effect since the $\theta$-term is not observable.

We expect this unobservability of the topological charge to also hold for other space-time dimensions.
Let us briefly inspect the case of the Schwinger model (quantum electrodynamics in 1+1-dimension) \cite{Schwinger:1962tn,Schwinger:1962tp} as an example.
The topological charge density in the Schwinger model is given by
\begin{eqnarray}
\frac{e}{\pi} \epsilon^{\mu \nu} \partial_\mu A_\nu
.
\end{eqnarray}
It is of course a total derivative, and the total topological charge becomes
\begin{eqnarray}
\frac{e}{\pi} \int dx \, A^1 (x) \Biggr|^{t=+\infty}_{t=-\infty}
=
\Delta n
.
\end{eqnarray}
Here $x$ is the one-dimensional space coordinate.
As for the case of Eq. (\ref{eq:topological_charge}), this may be nonzero if we have multiple windings of the gauge field while keeping the same value for the Wilson loop.
We see that $A^1$ is the only space-time degree of freedom left at $t=\pm \infty$, and it is without any doubt collinear to the gauge function since it is a one-dimensional quantity.
We therefore conclude that the topological charge and the $\theta$-term of the Schwinger model is also unphysical (we note that in 1+1-dimension, the topological charge is perturbation finite and not renormalized as long as the gauge boson is well behaving in the ultraviolet region \cite{Mastropietro:2007zza}.
This condition may be violated if there are contact interactions between fermions).
We may certainly derive the same conclusion for other topologically nontrivial gauge theories in other space-time dimensions.

In the next Section, we consolidate our argument by showing the unphysicalness of the topological charge using the WTI.

\section{BRST symmetry and Ward-Takahashi identity of topological charge\label{sec:BRST}}

In the previous section, we indirectly showed the unobservability of the topological charge by using the gauge function.
According to the BRST symmetry, processes involving unphysical gauge components must vanish at the level of observables.
One may then wonder how this is realized for the case of the topological charge.
Let us now try to justify it by using the WTI of the BRST symmetry.

The BRST transformations \cite{Becchi:1975nq,Tyutin:1975qk} of the fermion, gauge field, Faddeev-Popov ghost, anti-ghost, and Nakanishi-Lautrup (NL) field are given, respectively,  as follows:
\begin{eqnarray}
\delta_B \psi 
&=&
i\lambda 
g_s c_a ( t_a \psi )
,
\label{eq:BRS1}
\\
\delta_B A_a^\mu
&=&
\lambda
(
\partial^\mu c_a
+g_s f_{abc} A_b^\mu c_c
)
,
\label{eq:BRS2}
\\
\delta_B c_a
&=&
-\frac{1}{2}
\lambda
g_s f_{abc} c_b c_c
,
\label{eq:BRS3}
\\
\delta_B \bar c_a
&=&
i\lambda
B_a
,
\label{eq:BRS4}
\\
\delta_B B_a
&=&
0
,
\label{eq:BRS5}
\end{eqnarray}
with $\lambda$ the infinitesimal parameter, and satisfies the nilpotency $\delta_B^2 =0$.
The WTI for the general BRST quartet $(A,C ; \bar{C},B)$ is
\begin{equation}
\langle 0 | \, \{ Q_B , T [A (x) \bar { C} (y) ] \} \, | 0 \rangle
=
\langle 0 | T [ A (x) { B} (y)  ]\, | 0 \rangle
-i\langle 0 | T[ C (x) \bar{ C} (y)] \, | 0 \rangle
=0
.
\label{eq:generalWTI}
\end{equation}
where $Q_B$ is the generator of the BRST transformation.
From Eqs. (\ref{eq:BRS2})-(\ref{eq:BRS5}), we may match $A = A_a^\mu$, $C = D_\mu c_a$, $\bar C = \bar c_a$, and $B = B_a$, which is the well-known WTI for the BRST symmetry \cite{Kugo:1979gm}.

Let us now construct a WTI for the topological charge.
We use the idea of Ref. \cite{Kugo:1978nc} which started the discussion with the following topological current
\begin{equation}
K_\mu
\equiv
\frac{\alpha_s}{8\pi}
\epsilon_{\mu \nu \rho \sigma}
\Biggl[
A^\nu_a F_a^{\rho \sigma }
-\frac{g_s}{3}
f_{abc}
A^\nu_a A^\rho_b A_c^\sigma 
\Biggr]
=
\frac{\alpha_s}{4\pi}
\epsilon_{\mu \nu \rho \sigma}
\Biggl[
A^\nu_a \partial ^\rho A_a^{\sigma }
+\frac{g_s}{3}
f_{abc}
A^\nu_a A^\rho_b A_c^\sigma 
\Biggr]
.
\label{eq:topologicalcurrent}
\end{equation}
Its total divergence $\partial^\mu K_\mu$ is just the topological charge density operator (\ref{eq:topological_charge_density}).
If we impose $A=K^\mu$, the corresponding Faddeev-Popov ghost operator $C$ then becomes
\begin{equation}
C=
{\cal C}_\mu
=
[iQ_B , K_\mu]
=
\frac{\alpha_s}{4\pi}
\epsilon_{\mu \nu \rho \sigma}
(\partial^\nu c_a) \partial^\rho A_a^\sigma
.
\label{eq:topologicalghost}
\end{equation}
The derivation is given in Appendix \ref{sec:appendixBRST}.
The other BRST doublet counterpart $(\bar C , B)$ which fulfills the WTI (\ref{eq:generalWTI}) is arbitrary as long as $\delta_B \bar C = B$, but the physically meaningful operator for the topological charge is almost unique.
Here we choose 
\begin{equation}
\bar C
=
\bar {\cal C}_\mu
=
\frac{\alpha_s}{8\pi}
\epsilon_{\mu \nu \rho \sigma}
(\partial^{-2} \partial^\nu \bar c_a) 
g_s f_{abc} A_b^\rho A_c^\sigma
.
\label{eq:topologicalantighost}
\end{equation}
The BRST transform of $\bar {\cal C}_\mu$ yields the corresponding NL operator 
\begin{equation}
B
=
{\cal B}_\mu
=
\{ Q_B , \bar C_\mu \}
=
\frac{\alpha_s}{8\pi}
g_s f_{abc} 
\epsilon_{\mu \nu \rho \sigma}
\Bigl[ 
(\partial^{-2} \partial^\nu B_a) 
A_b^\rho A_c^\sigma
+
2(\partial^{-2} \partial^\nu \bar c_a) 
(\partial^\rho c_b )
A_c^\sigma
+ 
g_s f_{bde} 
(\partial^{-2} \partial^\nu \bar c_a) 
A_d^\rho A_e^\sigma
c_c 
\Bigr] 
.
\label{eq:topologicalNL}
\end{equation}
For the derivation, see Appendix \ref{sec:appendixBRST}.

With the above topological BRST quartet, we have the following WTI:
\begin{equation}
\langle 0 | \, \{ Q_B , T [K^\mu (x) \bar {\cal C}^\nu (y) ] \} \, | 0 \rangle
=
\langle 0 | T [ K^\mu (x) {\cal B}^\nu (y)  ]\, | 0 \rangle
-i\langle 0 | T[ {\cal C}^\mu (x) \bar{\cal C}^\nu (y)] \, | 0 \rangle
=0
.
\label{eq:topologicalWTI}
\end{equation}
By taking the four-divergences with respect to $x$ and $y$, we find
\begin{equation}
\langle 0 | T [ \partial_\mu K^\mu (x) \partial_\nu {\cal B}^\nu (y)  ]\, | 0 \rangle
=0
,
\label{eq:topologicalderivativeWTI}
\end{equation}
where the ghost term dropped due to the contraction of the four-derivative with the Levi-Civita tensor [see Eq. (\ref{eq:topologicalghost})].
Here we omitted the derivatives of the time step functions which are not relevant in further discussions.
After taking the limit $|x - y| \to \infty$, the WTI becomes
\begin{equation}
\sum_{| \Omega \rangle \ne | 0 \rangle}
\langle 0| \partial_\mu K^\mu (x) | \Omega \rangle \langle \Omega| \partial_\nu {\cal B}^\nu (y) | 0\rangle
=
0
.
\label{eq:topologicalchargeclusterdecomposition}
\end{equation}
It is evident that the cluster decomposed topological charge density is proportional to the topological charge, as $\langle 0| \partial_\mu K^\mu (x) | \Omega \rangle \propto \int d^4 x \, \partial_\mu K^\mu (x) $.
Here we denoted by $| \Omega \rangle$ vacua which belong to other topological sectors than that of $|0\rangle$.
We note that the transition between vacua with the same topology through the topological charge operator cancels.
Here we obtain a WTI which does not have a ghost term, which means that the gauge-NL term is intrinsically unphysical on its own.
The absence of ghost contribution can also be seen in the light-cone gauge or in quantum electrodynamics where the ghost modes decouple from the rest.

Let us now inspect the cluster decomposed topological NL operator $\langle \Omega|\partial_\nu {\cal B}^\nu (y) | 0\rangle$.
First, consider the general gauge fixed BRST invariant Lagrangian with the quartet fields
\begin{equation}
{\cal L}
=
\left[
-\frac{1}{4} (\partial^\mu A^\nu_a -\partial^\nu A^\mu_a)^2
-(\partial_\mu B_a ) A^\mu_a
+ \frac{\alpha_r}{2} B_a B_a
-i(\partial^\mu \bar c_a )( \partial_\mu c_a )
+g_s A^\mu_a \Gamma_{\mu a}
\right]
\varphi
,
\end{equation}
where $\alpha_r$, $\varphi$, and $\Gamma_{\mu a}$ denote the renormalized gauge fixing parameter, the sum of all possible gauge invariant (pseudo)scalar operators (including unity), and that of all color vector vertex, respectively.
By applying the variational principle to the gauge field $A^\mu_a$, we obtain the following equation of motion:
\begin{eqnarray}
\partial^2 A_a^\mu - \partial^\mu (\partial_\nu A_a^\nu ) 
&=& 
\partial^\mu B_a
-g_s \Gamma_a^\mu
\label{eq:GFEOM1}
.
\end{eqnarray}
Combining with Eq. (\ref{eq:topologicalNL}), we have
\begin{eqnarray}
\partial^\mu {\cal B}_\mu
&=&
\frac{\alpha_s}{8\pi}
g_s f_{abc} 
\epsilon_{\mu \nu \rho \sigma}
\partial^\mu
\Biggl[ 
\Bigl\{
A^\nu_a 
A_b^\rho A_c^\sigma
-
(\partial^{-2} \partial^\nu \partial_\alpha A^\alpha_a )
A_b^\rho A_c^\sigma
+
g_s
(\partial^{-2} \Gamma^\nu_a) 
A_b^\rho A_c^\sigma
\Bigr\}
\nonumber\\
&& \hspace{7em}
+
2
(\partial^{-2} \partial^\nu \bar c_a) 
(\partial^\rho c_b )
A_c^\sigma
+ 
g_s f_{bde} 
(\partial^{-2} \partial^\nu \bar c_a) 
A_d^\rho A_e^\sigma
c_c 
\Biggr] 
.
\ \ \ \ 
\label{eq:topologicalNLEOM}
\end{eqnarray}
We see that the first term yields the topological charge (\ref{eq:topological_charge}) after taking the vacuum matrix element, while the other terms have no effect since they are total divergences and also because these remaining terms will never generate the topological charge density via quantum corrections thanks to Adler-Bardeen theorem \cite{Adler:1969er,Adler:1972zd,Higashijima:1981km,Costa:1977pd,Fujikawa:1981rw,Lucchesi:1986sp,Anselmi:2014kja,Anselmi:2015sqa,Mastropietro:2020bhz}, so that they cancel the matrix elements between vacua with different winding numbers.
We also note that this expression does not depend on the gauge parameter $\alpha_r$.
We finally obtain from the WTI (\ref{eq:topologicalchargeclusterdecomposition}) 
\begin{eqnarray}
0
&=&
\sum_{| \Omega \rangle \ne | 0 \rangle}
\langle 0| 
\partial_\mu K^\mu (x) |\Omega  \rangle \langle \Omega | \partial_\nu {\cal B}^\nu (y) | 0\rangle
\nonumber\\
&\propto&
\sum_{| \Omega \rangle \ne | 0 \rangle}
\langle 0| 
F \tilde F (x) 
|\Omega \rangle \langle \Omega | 
F \tilde F (y) 
| 0\rangle
,
\label{eq:FFtilde_squared_amplitude}
\end{eqnarray}
This means that the topological charge of the vacuum is not observable.
Here the sum over topological sectors is responsible for the cancellation.
We note that this cancellation would not work if the initial and final state vacua are selected, but we do not have any means to control topological sectors, so the sum always run over all of them, and consequently the topological charge is unphysical.

We are interested in whether the topological charge of physical states may be probed or not.
The correlation of interest is then
\begin{equation}
\langle {\rm phys}' |
F\tilde F
| {\rm phys} \rangle
\equiv
\langle 0 |
F\tilde F (x) \phi (x')
| 0 \rangle
,
\end{equation}
where $\phi $ is the BRST singlet operator encoding the information of physical states.
We may repeat the same derivation as the vacuum case with this composite operator, and we arrive at the following relation:
\begin{eqnarray}
\sum_{| \Omega' \rangle
}
\langle 0| 
\phi (x') 
F \tilde F (x) 
|\Omega' \rangle \langle \Omega' | 
\partial_\nu {\cal B}^\nu (y)
\phi (y') | 0\rangle
&=&
0
.
\label{eq:FFtilde_obs_squared_amplitude}
\end{eqnarray}
This cluster decomposition is realized if the limits $|x-y| ,|x'-y| ,|x-y'| ,|x'-y'| \to \infty$ are taken.
We now have to inspect the sum over topological sectors $|\Omega' \rangle$.
If the operators at $x ,x'$ ($y,y'$) are not separated by an infinite distance (i.e. correlated), we may apply the operator product expansion (OPE) to $\phi (x') F \tilde F (x)$ ($F \tilde F (y) \phi (y')$) at the space-time coordinate $\frac{x+x'}{2}$ ($\frac{y + y'}{2}$).
According to Adler-Bardeen theorem \cite{Adler:1969er,Adler:1972zd,Higashijima:1981km,Costa:1977pd,Fujikawa:1981rw,Lucchesi:1986sp,Anselmi:2014kja,Anselmi:2015sqa,Mastropietro:2020bhz}, corrections to $F \tilde F $ cannot generate itself, so the OPE of $\phi (x') F \tilde F (x)$ ($F \tilde F (y) \phi (y')$) cannot generate $F \tilde F $ either, i.e. 
\begin{equation}
\phi (x') 
F \tilde F (x)
=
\sum_{O_i \ne F\tilde F }
C_i
O_i \biggl( \frac{x'+ x}{2} \biggr)
,
\label{eq:OPE1}
\end{equation}
with a finite distance between $x$ and $x'$.
This means that $\phi (x') F \tilde F (x)$ cannot change the winding number unless the two operators are uncorrelated by being infinitely separated so as to cluster decompose the topological charge density operator.
The same applies for $\partial_\nu {\cal B}^\nu$:
\begin{equation}
\partial_\nu {\cal B}^\nu (y) \phi (y') 
=
\sum_{O'_i \ne F\tilde F }
C'_i
O'_i \biggl( \frac{y'+ y}{2} \biggr)
.
\label{eq:OPE2}
\end{equation}
As we can see, the operator products lose the ability to change topological sectors due to Adler-Bardeen theorem for correlated operators with finite distances between $x$ and $x'$ ($y$ and $y'$).
When $x,x',y,y'$ are infinitely separated each other, however, the topological charge operators can be factorized:
\begin{eqnarray}
\sum_{| \Omega \rangle \ne | 0 \rangle}
\langle 0| \phi (x')  |0 \rangle \langle 0 | 
F \tilde F (x) 
|\Omega \rangle \langle \Omega | \partial_\nu {\cal B}^\nu (y) |0 \rangle \langle 0 | \phi (y') | 0\rangle
&&
\nonumber\\
\propto
\sum_{| \Omega \rangle \ne | 0 \rangle}
\langle 0| \phi (x')  |0 \rangle \langle 0 | 
F \tilde F (x) 
|\Omega \rangle \langle \Omega | 
F \tilde F (y) 
|0 \rangle \langle 0 | \phi (y') | 0\rangle
&=&
0
.
\label{eq:FFtilde_obs_squared_amplitude2}
\end{eqnarray}
The above expression contains Eq. (\ref{eq:FFtilde_squared_amplitude}) in a factorized form and it vanishes, which means that processes involving the topological charge are unphysical. 

If $\phi $ is a composite operator made of other $ F \tilde F $ operators, we may iteratively separate them by infinite distances to make them contribute to the topology change as
\begin{eqnarray}
\sum_{\Omega_1, \cdots , \Omega_{2n-1}  }
\langle 0| \psi (x')  |0 \rangle \langle 0 | 
F \tilde F (x_1) 
|\Omega_1 \rangle \langle \Omega_1 | 
F \tilde F (x_2) 
|\Omega_2 \rangle
\times
\cdots 
&&
\nonumber\\
\times
\langle \Omega_{2n-2} | 
F \tilde F (y_2) 
|\Omega_{2n-1} \rangle \langle \Omega_{2n-1} | 
F \tilde F (y_1) 
|0 \rangle \langle 0 | \psi (y') | 0\rangle
&=&
0
,
\end{eqnarray}
but the total amount nevertheless vanishes.
Again, this expression is a factorization containing Eq. (\ref{eq:FFtilde_squared_amplitude}) with the amplitude squared summed over all possible vacuum topologies.
This shows that the probability of realizing an arbitrary process with the operator $\phi$ vanishes for all cases admitting topology change with $F\tilde F$, which means that the topological charge does not affect physical observables for arbitrary power of $F\tilde F$. 

Particularly interesting targets are correlators of topological charge density
\begin{equation}
\langle 0 | T \{ F \tilde F (x_1) F \tilde F (x_2) \cdots F \tilde F (x_n) \} | 0 \rangle
.
\label{eq:topologicalcorrelator}
\end{equation}
From the above discussion, correlators of infinitely separated topological charge operators become
\begin{eqnarray}
\sum_{\Omega_1, \cdots , \Omega_{2n-1}  }
\langle 0 | 
F \tilde F (x_1) 
|\Omega_1 \rangle \langle \Omega_1 | 
F \tilde F (x_2) 
|\Omega_2 \rangle
\times
\cdots 
&&
\nonumber\\
\times
\langle \Omega_{2n-2} | 
F \tilde F (y_2) 
|\Omega_{2n-1} \rangle \langle \Omega_{2n-1} | 
F \tilde F (y_1) 
| 0\rangle
&=&
0
,
\label{eq:vacuumtopologychange}
\end{eqnarray}
which means that the correlations among topological charge density operators cannot probe the topology, no matter how many times it is changed.
We then arrive at the unobservability of the topological sectors of the vacuum, and also at the $\theta$-independence of the vacuum.
Of course, if the distances between coordinates of operators are all finite so as to correlate all operators, the correlators will have physical effect, but the transition of vacuum topology will be lost according to Adler-Bardeen theorem, as seen above.

\section{Inspection of the fermion and the chiral Ward-Takahashi identity\label{sec:fermion}}

We now introduce the fermions (quarks) with $N_f$ flavors and see the consistency of the chiral (or axial) WTI \cite{Bardeen:1969md}
\begin{equation}
\sum_{\psi}^{N_f}
\Bigl[
\partial^\mu (\bar \psi \gamma_\mu \gamma_5 \psi )
+2m_\psi
\bar \psi i\gamma_5 \psi
\Bigr]
=
-
N_f
F\tilde F
,
\label{eq:chiralWTI}
\end{equation}
with our findings.
We previously saw that the topological charge is unobservable.
There should then also be an unphysical contribution on the left-hand side of Eq. (\ref{eq:chiralWTI}).
From Atiyah-Singer's index theorem \cite{Atiyah:1963zz,Atiyah:1968mp,Atiyah:1968rj} which relates the zero-modes of the Dirac operator $D\hspace{-0.6em}/\, \equiv \partial \hspace{-0.5em}/\, -ig_s A \hspace{-0.55em}/\,_a t_a$ to the topological charge as
\begin{equation}
{\rm ind}(D\hspace{-0.55em}/\,)
=
-N_f 
\int d^4 x\,
F\tilde F
,
\label{eq:atiyah-singer}
\end{equation}
where ${\rm ind}(D\hspace{-0.55em}/\,)$ is the difference between the numbers of Dirac zero-modes with positive (right-handed) and negative (left-handed) chiralities, we may easily infer that the contribution from chiral Dirac zero-modes is the unphysical piece of Eq. (\ref{eq:chiralWTI}).

By removing this unobservable quark contribution from the left-hand side of Eq. (\ref{eq:chiralWTI}), we end up with the ``physical chiral WTI''
\begin{equation}
\sum_\psi^{N_f}
\Bigr[
\partial^\mu (\bar \psi \gamma_\mu \gamma_5 \psi )
+2m_\psi
\bar \psi i\gamma_5 \psi
\Bigr]_{\lambda \ne 0}
=
-N_f 
F\tilde F
\Bigr|_{\Delta n =0}
,
\label{eq:physicalchiralWTI}
\end{equation}
where the subscript of the left-hand side $\lambda \ne 0$ means that we removed the chiral Dirac zero-modes, and that of the right-hand side $\Delta n =0$ that we do not consider gauge configurations changing the winding number.
This separation of the physical fermionic modes from the topological charge and chiral Dirac zero-modes implies that the chiral phase rotation (= $U(1)_A$ transformation) of the fermion $\psi$ does not affect anymore the $\theta$-term and vice versa.
We also note that Eq. (\ref{eq:physicalchiralWTI}) still violates the $U(1)_A$ symmetry due to its right-hand side, but this violation is only local.
The space-time integration of the total divergent gluonic term vanishes so that the global $U(1)_A$ symmetry is conserved up to the fermion mass, without being affected by the topological charge.
The $U(1)_A$ symmetry will then suffer from spontaneous chiral symmetry breaking, just like the axial flavor nonsinglet charges.
This fact will actually drastically change the phenomenology, as explained in the following sections.

Let us be a little clearer with the local violation of the symmetry by the anomaly.
The most famous example is the decay of $\pi^0$ mesons to two photons.
This process shows that $U(1)_A$ is violated by the anomaly generated by combining the latter symmetry with two photonic $U(1)$ currents.
This violation however occurs only locally, at the vertex where $\pi^0$ decays. 
If the $U(1)_A$ current is averaged over the space-time (this corresponds to the case where $\pi^0$ has a vacuum expectation value), the anomaly will have no effect, since the $F_{\mu \nu} \tilde F^{\mu \nu}$ term of quantum electrodynamics is a total divergence.
If we apply this to QCD, we may find the same local $U(1)_A$ violation by replacing the photons by gluons, and $\pi^0$ by (the flavor singlet component of) $\eta'$.
We note that the irrelevance of the anomalous $U(1)_A$ violation is less pronounced because it is also explicitly broken by the quark mass.
For the case of $U(1)_{B+L}$ anomaly relevant in the SM, it is however much clearer.
Since we have no explicit $U(1)_{B+L}$ violation in the SM, this global symmetry is preserved.
The local violation of $U(1)_{B+L}$ via the anomaly may occur if we have particle analogues of $\pi^0$ or $\eta'$ which carry both the lepton ($L$) and baryon ($B$) numbers, but of course they do not exist in the SM.
Conversely, we may use the two-weak boson decays to search for new particles beyond the SM having $B+L$ in accelerator experiments.

\section{Symmetries of QCD and spontaneous breaking of $U(1)_A$\label{sec:QCDsym}}

Let us now discuss the consistency of the relevance of the $U(1)_A$ symmetry in the QCD Lagrangian, and its spontaneous breaking.
We first review the global symmetry of the quark sector of (massless) QCD \cite{Fukushima:2010bq,Tanizaki:2018wtg}.
The symmetry of the quarks at the level of the Lagrangian is
\begin{eqnarray}
G^{\rm (classical)}
&=&
U(N_f)_L \times U(N_f)_R
\nonumber\\
&=&
\frac{SU(N_f)_L \times SU(N_f)_R \times U(1)_L \times U(1)_R}{(\mathbb{Z}_{N_f})_L \times (\mathbb{Z}_{N_f})_R}
\nonumber\\
&=&
\frac{SU(N_f)_L \times SU(N_f)_R \times U(1)_V \times U(1)_A}{(\mathbb{Z}_{N_f})_L \times (\mathbb{Z}_{N_f})_R \times \mathbb{Z}_2}
.
\label{eq:quarkgroup}
\end{eqnarray}
The second equality was obtained using 
\begin{equation}
U(N_f) 
=
\frac{SU(N_f) \times U(1)}{\mathbb Z_{N_f}}
,
\end{equation}
where the discrete $\mathbb Z_{N_f}$ symmetry in the denominator is due to the following redundancy
\begin{equation}
e^{\frac{2\pi i}{N_f} }
=
e^{\frac{2\pi i}{N_f} {\rm diag}[1,\cdots , 1, 1-N_f] }
.
\end{equation}
The left-hand side of the above equation is the $U(1)$ rotation by the angle $\frac{2\pi i}{N_f}$ which is part of the discrete $\mathbb Z_{N_f}$ transformation, while the right-hand side is an element of the $SU(N_f)$ group (in $SU(3)$ group, this is the transformation generated by the Gell-Mann matrix $\lambda_8$).

The third equality of Eq. (\ref{eq:quarkgroup}) was derived using
\begin{equation}
U(1)_L \times U(1)_R
=
\frac{U(1)_V \times U(1)_A}{\mathbb Z_2}
.
\label{eq:U1LU1RU1VU1A}
\end{equation}
The explicit representations of the $U(1)_L$, $U(1)_R$, $U(1)_V$, and $U(1)_A$ are given by $e^{i\frac{1-\gamma_5}{2}\alpha_L}$, $e^{i\frac{1+\gamma_5}{2}\alpha_R}$, $e^{i\frac{\alpha_R+\alpha_L}{2}}$, and $e^{i\gamma_5 \frac{\alpha_R-\alpha_L}{2}}$, respectively, and we have a redundancy at the angle $\pi$ ($e^{i\pi}=e^{i\gamma_5 \pi}=-1$), corresponding to $\mathbb Z_2$ of the denominator \cite{Tanizaki:2018wtg}.

We may further transform Eq. (\ref{eq:quarkgroup}), but we have to be careful about whether the flavor number $N_f$ is even or odd.
When $N_f$ is odd, $SU(N_f)_V $, $ SU(N_f)_A $, $(\mathbb{Z}_{N_f})_V $, and $ (\mathbb{Z}_{N_f})_A $ have no elements corresponding to the multiplication of $-1$ like in Eq. (\ref{eq:U1LU1RU1VU1A}), so we have
\begin{eqnarray}
SU(N_f)_L \times SU(N_f)_R
&=&
SU(N_f)_V \times SU(N_f)_A
,
\\
(\mathbb{Z}_{N_f})_L \times (\mathbb{Z}_{N_f})_R
&=&
(\mathbb{Z}_{N_f})_V \times (\mathbb{Z}_{N_f})_A
.
\end{eqnarray}
When $N_f$ is even, however, $SU(N_f)_V $, $ SU(N_f)_A $, $(\mathbb{Z}_{N_f})_V $, and $ (\mathbb{Z}_{N_f})_A$ have the multiplication of $-1$ as a common element, so we have
\begin{eqnarray}
SU(N_f)_L \times SU(N_f)_R
&=&
\frac{SU(N_f)_V \times SU(N_f)_A}{\mathbb{Z}_2}
,
\\
(\mathbb{Z}_{N_f})_L \times (\mathbb{Z}_{N_f})_R
&=&
\frac{(\mathbb{Z}_{N_f})_V \times (\mathbb{Z}_{N_f})_A}{\mathbb{Z}_2}
.
\end{eqnarray}
In Eq. (\ref{eq:quarkgroup}), we have the former group divided by the latter, so $\mathbb{Z}_2$ is actually effectively irrelevant.
Eventually, the symmetry of the quark sector becomes
\begin{eqnarray}
G^{\rm (classical)}
&=&
\frac{SU(N_f)_V \times SU(N_f)_A \times U(1)_V \times U(1)_A}{(\mathbb{Z}_{N_f})_V \times (\mathbb{Z}_{N_f})_A \times \mathbb{Z}_2}
.
\label{eq:quarkgroup2}
\end{eqnarray}

From the conventional understanding, the $U(1)_A$ transformation does not hold at the quantum level, since the path integral measure is not invariant under it \cite{Fujikawa:1979ay,Fujikawa:1980eg}:
\begin{equation}
{\cal D} \bar \psi {\cal D} \psi \to^{\hspace{-0.75em} A} 
{\cal D} \bar \psi {\cal D} \psi 
e^{2i N_f \int d^4 x \, F \tilde F}
.
\label{eq:anomalymeasure}
\end{equation}
The integral of the exponent is the topological charge, and it may only be an integer, so that it was previously considered that the classical $U(1)_A$ symmetry becomes $(\mathbb{Z}_{2N_f})_A$ at the quantum level.
The symmetry of quarks expected so far was then
\begin{eqnarray}
G^{\rm (previous)}
&=&
\frac{SU(N_f)_L \times SU(N_f)_R \times U(1)_V \times (\mathbb{Z}_{2N_f})_A}{(\mathbb{Z}_{N_f})_L \times (\mathbb{Z}_{N_f})_R \times \mathbb{Z}_2}
.
\end{eqnarray}
In our work, however, we found that the topological charge is unphysical, so this explicit breaking is absent, and $U(1)_A$ still remains as a symmetry of the Lagrangian even in the quantum field theory, and Eq. (\ref{eq:quarkgroup2}) remains valid.

As a comment, the topological charge looks like a simple complex phase of a matrix element of quantum mechanics in looking at Eq. (\ref{eq:anomalymeasure}).
By recalling that a path integral is a correlator which has to be multiplied by its complex conjugate to get observable quantities, we may probably interpret the topological charge as an overall unobservable complex phase.

Now the $U(1)_A / \mathbb{Z}_2$ symmetry of the Lagrangian will suffer from the spontaneous breakdown together with the $SU(N_f)_A/(\mathbb{Z}_{N_f})_A$ one.
This may also be seen by adding mass terms to all quarks.
Here we note that we must divide $U(1)_A $ by $ \mathbb{Z}_2$ because  $U(1)_A $ contains a $ \mathbb{Z}_2$ part which is either a multiplication of $-1$ or an identity, of course preserved even if quarks are massive.
Formally, the need for the above chiral symmetry breaking may easily be shown using 't Hooft's anomaly matching \cite{thooftanomalymatchingconodition} by forming triangle graphs with other vector symmetries such as $SU(N_f)_V$ (note that the anomaly matching argument also allows massless baryons and keeps the $U(1)_A$ symmetry, but this scenario is forbidden by Vafa-Witten theorem, given that the lightest quark is massive, which is now almost established from lattice simulations \cite{Alexandrou:2020bkd}).
The final symmetry of hadron physics should then be
\begin{eqnarray}
G^{\rm (hadron)}
&=&
\frac{SU(N_f)_V \times U(1)_B }{(\mathbb{Z}_{N_f})_V}
,
\label{eq:hadrongroup}
\end{eqnarray}
where we promoted the $U(1)_V$ symmetry of the quark to $U(1)_B = U(1)_V/\mathbb{Z}_{N_c}$ due to the confinement of quarks into baryons \cite{Tanizaki:2018wtg}.
The symmetry of Eq. (\ref{eq:hadrongroup}) is just the one used in many successful models of hadron physics such as the quark model or in chiral perturbation theory.
Remarkably, this result also agrees with the symmetry of the skyrmion which was introduced as a topological object of pion field to describe the nucleon \cite{Skyrme:1961vq,Skyrme:1962vh,Witten:1979kh,Adkins:1983ya} and recently shown that its symmetry matches that of QCD, even by assuming the explicit breaking of $U(1)_A$ into $(\mathbb{Z}_{2N_f})_A$ \cite{Tanizaki:2018wtg}.

Since the global $U(1)_A$ symmetry is broken by the dynamics of QCD, there should be a proper NG boson.
As another important point, it should be restored above some critical temperature, since $U(1)_A$ is now a physical symmetry which broke down spontaneously.
This is actually strongly suggested by recent lattice results \cite{Cossu:2013uua,Chiu:2013wwa,Brandt:2016daq,Ishikawa:2017nwl,Bazavov:2019www,Aoki:2020noz}, as seen in Sec. \ref{sec:lattice}.
We therefore do not see any inconsistency in QCD even by admitting the spontaneous symmetry breaking of $U(1)_A$.

\section{Consistency with the cluster decomposition (Strocchi's theorem)\label{sec:strocchi}}

In quantum field theory, it is known that the correlation between operators tends to zero in the limit of large spacelike separation between them.
For two operators without vacuum expectation values $A_1$ and $A_2$, we have 
\begin{equation}
\langle 0 | A_1 (x) A_2 (y) |0 \rangle
\hspace{1.em} 
\to
\hspace{-1.5em} 
_{{}_{\tiny |x-y|\to \infty }}
\hspace{+1.em}
\langle 0| A_1 (x) | 0 \rangle \langle 0 | A_2 (y) |0 \rangle
= 0
.
\label{eq:cluster_decomposition}
\end{equation}
This is due to the cluster decomposition property which is a general feature of local quantum field theory.
Here fields have to be defined without vacuum expectation value, since this will be in conflict with the locality.
More precisely, it holds under the assumption of \cite{Araki:1962zhd}
\begin{itemize}
\item
translation invariance,

\item
local commutativity,

\item
uniqueness of the vacuum.

\end{itemize}

Naively, one might find inconsistencies for nonabelian gauge theories where colored particles are confined, since the confinement is suggesting a correlation becoming larger and larger at long distance.
This problem may be resolved by extending the state vector space to the indefinite metric and reinterpreting the unphysicalness.
For this case, Strocchi stated the following theorem \cite{Strocchi:1975xz,Strocchi:1978ci,Lowdon:2015fig}:
assuming the same conditions as above, a two-point correlator between two operators $B_1$ and $B_2$ obeys
\begin{eqnarray}
&&| \langle 0| B_1 (x) B_2 (y) |0 \rangle
-\langle 0| B_1 (x) |0 \rangle \langle 0| B_2 (y) |0 \rangle |
\nonumber\\
&&
\leq
\left\{
\begin{array}{ll}
C [x-y]^{-3/2 + 2N} e^{-M [x-y]} (1+ |E_x-E_y| /[x-y])
& ({\rm spectrum\, with\, mass\, gap\,} M)
\cr
C' [x-y]^{-2 + 2N} (1+ |E_x-E_y| /[x-y]^2)
& ({\rm spectrum\, without\, mass\, gap})
\cr
\end{array}
\right.
,
\end{eqnarray}
where $[x-y] \approx$ $|${\bf x }$-$ {\bf y}$|$ when $x$ and $y$ are separated with a large spacelike distance (for the precise definition, see Ref. \cite{Araki:1962zhd}). 
We see that in the case without mass gap, the cluster decomposition may not hold.
This is actually the case of nonabelian gauge theory where massless particles with indefinite metric, such as Faddeev-Popov ghosts, are present.
Indeed, the quartet mechanism and the above cluster decomposition property are consistent, since states involving members of BRST quartet have indefinite metric.
Such states may not obey the cluster decomposition, but this is not a problem since they cannot be observed at large spacelike separation due to the confinement \cite{Kugo:1979gm}.
BRST singlet states of course obey the cluster decomposition principle, and they are observable.

Let us now inspect the consistency of the above discussion for the case of the topological charge.
We just found that the topological charge (\ref{eq:topological_charge}) obeys the WTIs (\ref{eq:FFtilde_squared_amplitude}) and (\ref{eq:FFtilde_obs_squared_amplitude2}).
These are sums of amplitudes squared over all topological sectors but the total vanishes, so the indefinite metric should be working in this cancellation.
It is therefore not a problem even if individual terms of Eqs. (\ref{eq:FFtilde_squared_amplitude}) and (\ref{eq:FFtilde_obs_squared_amplitude2}) do not respect the cluster decomposition.
Moreover, an isolated topological charge changes the winding number of the vacuum gauge field, which does not fulfill the assumption of the uniqueness of the vacuum.
The latter condition is outside Strocchi's assumptions.
We therefore do not see any explicit inconsistency in this discussion.

Regarding fermions, we do not have any inconsistent points either, because the same cancellation as Eqs. (\ref{eq:FFtilde_squared_amplitude}) and (\ref{eq:FFtilde_obs_squared_amplitude2}) will happen for the chiral Dirac zero-modes thanks to Atiyah-Singer theorem (\ref{eq:atiyah-singer}).
In this context, we also note that the physical part of the singlet axial charge [obeying the ``physical'' anomalous $U(1)_A$ WTI (\ref{eq:physicalchiralWTI})] is detached from the chiral Dirac zero-modes, and respects the standard cluster decomposition principle (\ref{eq:cluster_decomposition}).

\section{Impact of the unobservability of Dirac zero-modes on phenomenology\label{sec:zero-modes}}

It was so far believed that the symmetry explicitly broken by the anomaly generates the 't Hooft vertex, an effective contact multi-fermion interaction having a determinant form \cite{tHooft:1976rip,tHooft:1976snw}.
For instance, in QCD with broken $U(1)_A$, we have the following interaction (the so-called Kobayashi-Maskawa-'t Hooft interaction \cite{Kobayashi:1970ji,Kobayashi:1971qz,Hatsuda:1994pi})
\begin{equation}
{\cal L}_{\rm KMT}
=
C \bar u_R u_L \bar d_R d_L \bar s_R s_L + {\rm h.c.}
,
\label{eq:KMT}
\end{equation}
where $C$ is a phenomenological parameter which may be obtained from the analysis of the instanton at low energy.
This three-quark interaction is $SU(3)_L \times SU(3)_R$ invariant, and it is generated by the chiral Dirac zero-modes \cite{tHooft:1976rip,tHooft:1976snw}.
From the discussion of Sec. \ref{sec:fermion}, we can affirm that observable interactions violating the $U(1)_A$ symmetry via chiral Dirac zero-modes cannot happen, since they belong to the unphysical space of the gauge theory.
This is consistent with the fact that no direct effects of the 't Hooft vertex have been observed in accelerator experiments \cite{Khoze:2019jta,Khoze:2020tpp,Khoze:2020paj} as well as the difficulty to set an evidence for the chiral magnetic effect \cite{Kharzeev:2007jp,Fukushima:2008xe,Kharzeev:2015znc} which was expected to be an excellent probe of the chiral chemical potential due to the topological charge.
We also point out that our result will upset the scenario where the dynamical chiral symmetry breaking is induced by the strong attraction of the instanton-induced multiquark interaction of Eq. (\ref{eq:KMT}) \cite{Hatsuda:1994pi,Schafer:1996wv}, as well as other phenomenological models augmented by it \cite{Oka:1989ud,Oka:2000wj}.
Here we make a remark that Eq. (\ref{eq:KMT}), if the origin is not due to the topology but rather due to the dynamical chiral symmetry breaking of QCD (or by explicit breaking due to current quark masses), is totally allowed in low energy effective models of QCD, and results of previous works will not drastically change as long as the interaction couplings are fitted to reproduce experimental data.

However, it is difficult to believe that the KMT interaction (\ref{eq:KMT}) persists at and above the critical temperature and density.
Indeed, recent lattice calculations are suggesting that the violation of $U(1)_A$ disappears above the critical temperature $T_c$ in QCD \cite{Cossu:2013uua,Chiu:2013wwa,Brandt:2016daq,Ishikawa:2017nwl,Bazavov:2019www,Aoki:2020noz} (see also Sec. \ref{sec:lattice}).
As another interesting point, it is phenomenologically known that Eq. (\ref{eq:KMT}) mixes the chiral and the diquark condensates and that the transition of QCD at low temperature and finite density becomes a crossover due to this \cite{Hatsuda:2006ps,Abuki:2010jq}.
This crossover nature is actually an attractive candidate which may explain the stability of the neutron star by rendering the equation of state of the quark and baryonic matter stiffer \cite{Baym:2017whm}.
The unphysicalness of the KMT interaction is then important in the sense that it will force this scenario to be modified.
We however note that the KMT interaction is not the only way to resolve the problem of the neutron star stability, so more detailed studies are definitely needed in the future, although we do not discuss them here.

Another very important example of multi-fermion interaction arising from chiral Dirac zero-modes is the $U(1)_{B+L}$ violation of the electroweak theory.
This gives rise to the baryon and lepton number violations, and it was so far the most promising process which could explain the baryogenesis \cite{Sakharov:1967dj,Riotto:1999yt} via the sphaleron process \cite{Manton:1983nd,Klinkhamer:1984di,Fukugita:1986hr} occurring at high temperature in the early universe.
Since writing the explicit form of the Lagrangian of the interaction with 12 fermion legs is somewhat tedious, here we just give an example of processes which are induced by it:
\begin{equation}
u + d 
\to
\bar d + \bar s + 2\bar c +3\bar t + e^+ + \mu^+ + \tau^+
.
\end{equation}
Here it is to be noted that all particles are $SU(2)_L$ doublets.
This interaction is induced by the anomaly between the baryon/lepton number current and the electroweak gauge ones which are chiral.
Again, this zero-mode induced process is physically forbidden, and this leads us to think of particle physics models within which explicit baryon number violating interactions are contained \cite{Georgi:1974sy,Yoshimura:1978ex,Affleck:1984fy,Dine:2003ax,Dorsner:2016wpm}, if one wishes to explain the baryon abundance around us.
Obviously, the SM is unable to realize the matter abundant Universe, independently of the amount of CP violation.

We note that the asymmetry required to explain the matter abundance is only for baryons and not for leptons, because the electrons around us might have been created after the neutron beta decays.
The current constraint on the lepton number asymmetry is not strong, and its upper limit is only at the level of $O(10^{-3})$ \cite{Oldengott:2017tzj,Pitrou:2018cgg}, while the baryon number asymmetry is $O(10^{-10})$.
This means that even a radical scenario where only the baryon number violating interactions were relevant in the early Universe and where the lepton number violation is absent is totally allowed.
Currently, the survey of extremely metal-poor galaxies is suggesting a large lepton number \cite{Matsumoto:2022tlr}, which is the other extremal case.
We note that the generation of this large lepton number asymmetry is allowed to happen before the electroweak phase transition, since the sphaleron process is unphysical and no charge transfer to the baryonic sector occurs.

\section{Topological susceptibility, $\eta'$ and consistency with large $N_c$\label{sec:etaprime}}

\begin{figure}[htb]
\begin{center}
\includegraphics[width=10cm]{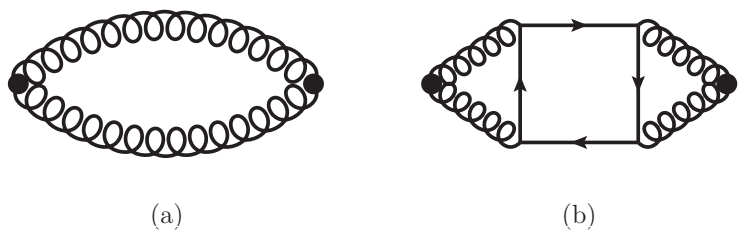}
\caption{\label{fig:eta_prime_gluon}
The leading (perturbative) contribution to the topological susceptibility (\ref{eq:topological_susceptibility}) in the $1/N_c$ expansion (a), and a subleading quark loop contribution (b).
}
\end{center}
\end{figure}

The large $N_c$ analysis is a framework which is qualitatively successful in the description of QCD processes \cite{tHooft:1973alw}.
In the case of the $U(1)_A$ problem, however, there was an apparent inconsistency when massless quarks are introduced.
Let us first review the arguments of previous works which analyzed the $U(1)_A$ problem using the $1/N_c$ expansion \cite{Witten:1979vv,Veneziano:1979ec,Veneziano:1980xs}.
Their point was to inspect the vacuum energy dependence on the vacuum angle
\begin{equation}
E(\theta )
=
E_0 +\sum_n c_n \theta^n
,
\end{equation}
whose Taylor coefficients were assumed to be given by the correlator
\begin{equation}
c_n
=
\frac{d^n E}{d \theta^n} \Biggl|_{\theta =0}
=
\int d^4 x_1 d^4 x_2 \cdots d^4 x_n 
\langle 0|
T\{
F\tilde F (x_1)
F\tilde F (x_2)
\cdots
F\tilde F (x_n)
\}
|0\rangle
.
\label{eq:theta_vacuum_energy_Taylor}
\end{equation}
Let us inspect the second order correlator with momentum dependence (the so-called ``topological susceptibility''), defined as
\begin{equation}
U(k)
=
\frac{1}{N_c^2}
\int d^4 x\, e^{ik\cdot x}
\langle 0|
F\tilde F (x)
F\tilde F (0)
|0\rangle
.
\label{eq:topological_susceptibility}
\end{equation}
The leading contribution of this correlator in the $1/N_c$ expansion is depicted in Fig. \ref{fig:eta_prime_gluon} (a), and it is an $O(1)$ effect (the factor $1/N_c^2$ has been inserted for normalizing it).
Then, what if massless quarks are introduced?
Before our discussion, the understanding was that massless quarks permit the chiral rotation to unphysicalize the $\theta$-term, and previous works argued that the addition of massless quarks in the theory must cancel the vacuum energy dependence, i.e. Eqs. (\ref{eq:theta_vacuum_energy_Taylor}), (\ref{eq:topological_susceptibility}).
However, the quark effect is subleading in the $1/N_c$ expansion [see Fig. \ref{fig:eta_prime_gluon} (b)] and it cannot naively remove the leading order purely gluonic contribution, so there was a paradox to resolve.
References \cite{Witten:1979vv,Veneziano:1979ec,Veneziano:1980xs} solved this problem with the following arguments.
By assuming that the topological susceptibility (\ref{eq:topological_susceptibility}) does not vanish in the limit of vanishing momentum, it should be given as
\begin{equation}
U(k)
=
\sum_{i={\rm glueballs}}
\frac{a_i}{k^2 - M_i^2}
+
\sum_{j={\rm mesons}}
\frac{1}{N_c} \cdot
\frac{c_j }{k^2 - m_j^2}
,
\end{equation}
where $a_i$ and $c_j$ are $O(1)$ factors in the $1/N_c$ expansion.
The first term is generated by the pure glue dynamics, while the second one is the contribution from all diagrams with one fermion loop inserted.
For the general momentum $k$, the two terms of $U(k)$ do not cancel.
However, for the vacuum ($k=0$), the cancellation may happen if there is a meson $\eta'$ with the mass
\begin{equation}
m_{\eta'}^2
=
\frac{1}{N_c} \cdot
\frac{c_{\eta'} }{ \sum_{i={\rm glueballs}}
\frac{a_i}{ - M_i^2}}
.
\label{eq:eta'mass}
\end{equation}
This results in an $N_c$-dependent formula for the mass of the 9th ``would-be NG boson'' $\eta'$, required if the chiral rotation erases the $\theta$-term.

Let us now see how the above discussion is modified after our findings.
We saw in Sec. \ref{sec:BRST} that the vacuum has no $\theta$-dependence since the topological sector and the vacuum angle $\theta$ are both unphysical.
We therefore do not need any cancellation between the leading $O(1)$ gluonic contribution and the subleading $O(N_c^{-1})$ quark loop effects as regards topological correlators.
As another point, we obtained the important result that $\theta$ is not affected by chiral rotations.
The cancellation then never happens and the constraint (\ref{eq:eta'mass}) becomes irrelevant.
The vacuum energy is free from both $\theta$ and chiral rotation.
This then resolves the paradox of $1/N_c$ expansion.

Let us now inspect the fate of $\eta'$ which is believed to be generated by the topological charge density operator (\ref{eq:topological_charge_density}).
In the current qualitative understanding \cite{Bass:2018xmz,Gan:2020aco}, $\eta'$ is generated by the mixing between the flavor $SU(3)$ singlet quark-antiquark state and that generated by the $\lambda_8$ generator of flavor $SU(3)$ due to the strange quark mass.
The singlet state also receives additional contribution from the topological susceptibility (\ref{eq:topological_susceptibility}).
The $\eta'$ meson is the heaviest state of this mixing with $m_{\eta'}\approx 960$ MeV (the lightest one is $\eta$, with $m_{\eta}\approx 550$ MeV).
We note that $\eta'$ results from a large mixing between the flavor singlet and nonsinglet states.
The contribution of the topological susceptibility (\ref{eq:topological_susceptibility}) suggests that $\eta'$ also mixes with the pseudoscalar glueball.
However, the pseudoscalar glueball is much heavier than $\eta'$ ($> 2$ GeV) \cite{Morningstar:1999rf,Sun:2017ipk}, so $m_{\eta'}$ is not much modified even if we account for the mixing with glueballs (also confirmed in lattice QCD \cite{Jiang:2022ffl}).

Our conclusion was that the $U(1)_A$ symmetry is not explicitly broken by the chiral anomaly but only spontaneously by the dynamics of QCD, so $m_{\eta'}$ (and $m_\eta$) must tend to zero as the masses of the up, down, and strange quarks become small.
The topological susceptibility (\ref{eq:topological_susceptibility}) which appears as a mass insertion in the flavor singlet channel must then vanish in this same limit.
When the current quarks are massive, the quark components of $\eta'$ (and $\eta$) are massive as well, so the topological susceptibility in the intermediate state will have an energy-momentum inflow.
From this fact, we may infer that the topological susceptibility has a momentum dependence, so that its contribution vanishes in the massless limit of quarks while it sufficiently increases the mass of $\eta'$ (and $\eta$) in real QCD due to the massive quarks in the Lagrangian.
Let us show that this is indeed the case.
After Fourier transforming Eq. (\ref{eq:topological_susceptibility}), we obtain
\begin{equation}
U(k)
=
\frac{1}{N_c^2}
\sum_h
\langle 0 | F \tilde F |h (k) \rangle \frac{i}{k^2 -m_h^2} \langle h (k) | F \tilde F | 0 \rangle
,
\label{eq:topologicalsusceptibilitymomentum}
\end{equation}
where $h$ is the hadronic intermediate state, and its summation may also run over the continuum.
Since the topological charge is a total divergence and the nonrenormalizability of the chiral anomaly conserves this property, $\langle 0 | F \tilde F |h (k) \rangle $ tends to zero in the limit $k^2\to 0$.
The correlator (\ref{eq:topologicalsusceptibilitymomentum}) safely tends to zero with the energy-momentum if all gluonic hadrons are massive, but it
may not vanish at $k^2 = 0$ if the lightest state is a massless hadronic boson.
In this regard, we point out that this massless hadron with zero momentum $k^\mu =0$ (which is obtain by Lorentz transforming arbitrary energy-momenta fulfilling $k^2=0$) is a constant field over the whole space-time, and the topological charge matrix element which creates $h$ is proportional to the vacuum matrix element $\langle 0 |F\tilde F | h (k^\mu=0) \rangle \propto \langle 0 |F\tilde F | \Omega \rangle$, so the topological susceptibility is either zero or unphysical with zero inflowing energy-momentum.
We therefore conclude that $\eta$ and $\eta'$ are both pseudo-NG bosons of QCD.

\section{Consistency with phenomenological instanton models\label{sec:instanton}}

Let us first review the physics of instantons.
The Euclidean Yang-Mills action may be rewritten as
\begin{equation}
S_{\rm YM}
=
\frac{1}{4}
\int d^4 x \, F_{\mu \nu}^a F_{\mu \nu}^a
=
\frac{1}{4}
\int d^4 x 
\Biggl[
\pm F_{\mu \nu}^a \tilde F_{\mu \nu}^a
+\frac{1}{2} \Bigl( F_{\mu \nu}^a \mp \tilde F_{\mu \nu}^a \Bigr)^2
\Biggr]
.
\end{equation}
The first term, the topological charge, is invariant, so we see that the action is minimal when the self-(anti)dual relation 
\begin{equation}
\tilde F_a^{\mu \nu}
=
\pm
F_a^{\mu \nu}
,
\label{eq:selfdual}
\end{equation}
is realized.
It was then believed that classical configurations obeying the above relation are the most frequently realized among all quantum processes, and tunnelings between different topological sectors were thought to happen with the probability
\begin{equation}
P
\sim
e^{-\frac{2\pi}{\alpha_s}\Delta n} 
.
\label{eq:instantonprobability}
\end{equation}
A solution satisfying Eq. (\ref{eq:selfdual}), the instanton, 
\begin{equation}
A_\mu^{{\rm (cl)} a}
=
2 \eta_{a\mu \nu} \frac{x^\nu}{x^2 +\rho^2}
,
\end{equation}
where $\eta = \epsilon_{a \mu \nu}\ ({\rm for }\, \mu , \nu = 1,2,3), -\delta_{a \mu}\ ({\rm for }\, \nu =4), \delta_{a \nu}\ ({\rm for }\, \mu =4)$, is known so far \cite{Belavin:1975fg}.
Here $\rho$ is an arbitrary parameter.
Lattice calculations also support the dominance of the instantons in gauge configurations \cite{Chu:1994vi}.
However, we saw in our discussion that the topological charge is unphysical.
The instanton, satisfying Eq. (\ref{eq:selfdual}), also has a topological charge, so this immediately leads to claim that it is unphysical.
Another remarkable point is that the exponent of the tunneling probability (\ref{eq:instantonprobability}) becomes unphysical (effectively, the exponent vanishes), so that all transitions between arbitrary topological sectors are realized with the same probability, and the transition effect may be factorized in the path integral at least for the classical contribution.
The number of instantons which make tunnelings is therefore irrelevant since it only represents an overall factor, and this is totally consistent with their unphysicalness.
It is interesting to note that the renormalization group evolution of $\alpha_s$ may be extracted from higher order loop corrections to the tunneling rate (\ref{eq:instantonprobability}) generated by the quantum fluctuation around the classical configuration $A_a^{{\rm (cl)}\mu} + \delta A_a^\mu$ \cite{tHooft:1976snw,Schafer:1996wv}.
This means that, while the vacuum tunneling due to the instanton has no physical effect, calculational techniques using it are still keeping practical importance.

Let us now see how the interpretation of phenomenological models involving instantons are affected by the result of our discussion.
A successful model using the instantons is the interacting instanton liquid model \cite{Shuryak:1981ff,Diakonov:1983hh,Gross:1980br}, where the gauge configuration calculated using variational principle is simplified as a superposition of instantons and anti-instantons as trial functions.
They are distributed in space-time according to phenomenological interactions between them.
They have spreads ($\rho$) of the order of the hadron size, so as long as there are no fluctuations with scales smaller than that of hadrons, it is naturally expected that the effects of hadron level interactions with ranges larger than $1/\Lambda_{\rm QCD}$ can be reproduced.
Here let us make the important remark that non-isolated multi-(anti)instanton configurations are not exact solutions of the self-(anti)duality equation (\ref{eq:selfdual}) due to the non-linearity.
Multi-(anti)instanton processes therefore involve topology nonchanging contributions, so gauge configurations having more than two (correlated) instantons can be physical.
With a sufficient number of (anti)instantons, it is then possible to fit physical gauge configurations.

In phenomenological models using instantons, important nonperturbative properties of quarks such as the chiral symmetry breaking are triggered by the 't Hooft vertex [see Eq. (\ref{eq:KMT})].
As seen in Sec. \ref{sec:fermion}, 't Hooft vertices are generated by chiral Dirac zero-modes and are thus unphysical.
Respecting our conclusion, they will have to be removed in those models.
This change will actually force us to correct our major understanding concerning the relation between the topological charge and nonperturbative aspects related to quarks.
We also note that phenomenological models combining instanton models with other conventional models of hadron physics such as the quark model \cite{Oka:1989ud,Oka:2000wj} and others \cite{Hatsuda:1994pi,Schafer:1996wv} have to be treated with caution regarding potential double countings.

\section{Fate of the axion mechanism\label{sec:axion}}

Let us now see the phenomenological consequence of the axion mechanism of Peccei and Quinn \cite{Peccei:1977hh}.
To review it in a few lines, the axion is a (pseudo)scalar particle which couples to the QCD topological term (\ref{eq:topological_charge_density}) through the interaction
\begin{equation}
{\cal L}_{a}
=
N_f \frac{a }{f_a} F\tilde F
,
\label{eq:axionlagrangian}
\end{equation}
with some decay constant $f_a$ which has the energy scale where the axion $a$ has been generated, and it was introduced to unphysicalize the $\theta$-term.
The cancellation of the effect of the $\theta$-term was expected to occur through the effective Lagrangian of the axion
\begin{equation}
{\cal L}_{a}^{\rm (eff)}
=
\frac{1}{2}
\partial^\mu a \partial_\mu a
- K_1 \Biggl( \theta +\frac{a }{f_a} \Biggr)
- \frac{1}{2}K_2 \Biggl( \theta +\frac{a }{f_a} \Biggr)^2
+ \cdots
.
\label{eq:axioneffectivelagrangian}
\end{equation}
If we do not consider the linear term (with $K_1$), this potential is minimized at $\theta = -\frac{\langle a \rangle}{f_a}$, and the effective topological operator $F\tilde F $ vanishes.
If the topological charge was physically observable, the Strong CP problem could have been resolved with this dynamical mechanism.

However, we saw that the topological charge is strictly unphysical.
What would then happen if we couple the axion to the strong interaction as in Eq. (\ref{eq:axionlagrangian})?
Actually, Eq. (\ref{eq:axionlagrangian}) is still physical, but the contribution from the vacuum expectation value of the axion to Eq. (\ref{eq:axionlagrangian}) becomes unphysical, since $\frac{\langle a \rangle }{f_a}F\tilde F $ is proportional to the topological charge density.
We note that the coefficient $K_2$ of the quadratic term of Eq. (\ref{eq:axioneffectivelagrangian}) is the topological susceptibility [see Eq. (\ref{eq:topological_susceptibility})].
This means that the axion cannot obtain mass from null, but its mass may be modified if it has a bare mass, like $\eta'$ and $\eta$.

Let us also comment on the linear term of Eq. (\ref{eq:axioneffectivelagrangian}) (with $K_1$).
This term was supposed to be generated by other CP violating quark-gluon level operators $O_{CP}$ such as the quark chromo-EDM, and it might have shifted the minimum of the axion potential.
Before our discussion came, this shift was believed to generate an ``induced $\theta$-term`` which cannot be erased even under the axion mechanism \cite{Shifman:1979if}.
Its observable effect was (supposed to be) suppressed by the scale of new physics relevant in $O_{CP}$.
After our discussion, this induced $\theta$-term will of course become unphysical.
Since the linear coupling is given by
\begin{equation}
K_1 =
-i
N_f
\lim_{k \to 0}
\int d^4 x\, e^{ik\cdot x}
\langle 
T\{
F\tilde F (x)
O_{CP}(0)
\}
\rangle
,
\label{eq:linearaxioncoupling}
\end{equation}
it actually cancels due to the zero inflowing energy-momentum to the topological charge density operator.
The $K_1$ term therefore vanishes.

The change in the physics of instantons and topological susceptibility might have an important impact on the axion dark matter scenario \cite{Preskill:1982cy,Abbott:1982af}, since the axions are coupled to the topological charge density of QCD.
Needless to say, this is of course only possible if at least one new light scalar field beyond the SM coupled to QCD via the interaction of Eq. (\ref{eq:axionlagrangian}) exists.
To evaluate the relic density of axion dark matter, we need to quantify the axion mass dependence on the temperature which is possible by evaluating the topological susceptibility using instanton models \cite{Turner:1985si,Wantz:2009it} or lattice QCD \cite{Berkowitz:2015aua}.
The instanton liquid models are predicting the following temperature dependence for the axion mass \cite{Wantz:2009it}:
\begin{equation}
m_a (T)^2
\approx
\alpha
\frac{\Lambda^4}{f_a^2} \left( \frac{T}{\Lambda} \right)^{-n}
,
\end{equation}
where $\Lambda = 400$ MeV, $\alpha = 1.68 \times 10^{-7}$, and $n=6.68$.
We note that the instanton liquid model is allowed to be physical, as seen in the previous Section.
The growth of the axion mass while the temperature falls is indeed important to generate light cold dark matter (the so-called ``misalignment mechanism'').
In this regard, it is not clear to us how the physical configurations with instantons correlated each other contribute to the generation of axion mass.
This point will have to be carefully discussed in the future since the evolution of the mass of axions in the temperature will significantly affect their production.

\section{Phenomenological constraints on CP violation beyond the standard model\label{sec:EDM}}

\begin{figure}[htb]
\begin{center}
\includegraphics[width=12cm]{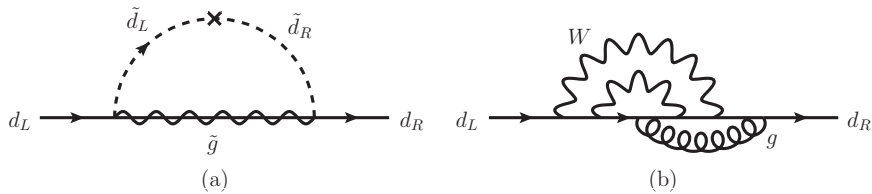}
\caption{\label{fig:BSMcpoddmass}
Examples of loop level diagrams contributing to the CP-odd mass of the down-quark in the minimal supersymmetric SM (a) \cite{Ellis:1982tk,Pospelov:2005pr} and in the SM (b) \cite{Ellis:1978hq,Khriplovich:1985jr}.
In Fig. (a), the loop particles include a gluino $\tilde g$ and two $d$-squark fields ($\tilde d_{L,R}$), which may have CP violating transition between them.
The internal solid lines of Fig. (b) must each have different flavors than $d$ to be CP violating.
These contributions are physical if so is the $\theta$-term, but now that the latter is unphysical, the CP phases of these processes may be removed after the redefinition of quark fields.
}
\end{center}
\end{figure}

One of the most important phenomenological consequences of our discussion after the resolution of the Strong CP problem is the modification of the constraints on the CP phases of new physics beyond the SM.
We saw that the topological contribution to the chiral WTI (\ref{eq:chiralWTI}) becomes irrelevant.
Before our discussion, it was believed that the $\theta$-term was relevant and that the difference between $\theta$ and the CP violating phase of the quark mass matrix (the so-called $\bar \theta$) was physical.
The axion mechanism \cite{Peccei:1977hh} could unphysicalize this $\bar \theta$, so that the CP violation of the new physics contributing to the CP-odd quark mass
\begin{equation}
{\cal L}_{\rm odd}
=
-m_{\rm odd} \bar \psi i \gamma_5 \psi
,
\label{eq:CP-odd_mass}
\end{equation}
was completely neutralized and constraints from EDM experiments were avoided.
However, now that the $\theta$-term is irrelevant, the CP-odd quark mass may be rotated away by redefining quark fields.
Indeed, the $U(1)_A$ transformation, which is now physical according to Sec. \ref{sec:fermion}, removes the CP phases of the quark mass terms generated by the SM as well as by new physics beyond it (see Fig. \ref{fig:BSMcpoddmass}), and we are now free of the CP-odd quark mass without using the axion mechanism.

Let us see how much the constraint from the neutron EDM experiment is relaxed after our discussion, with the supersymmetric model as an example.
Using the naive dimensional analysis, the contribution of the CP-odd quark mass of Fig. \ref{fig:BSMcpoddmass} (a) to the neutron EDM, when the $\theta$-term is physical, is \cite{Ellis:1982tk,Pospelov:2005pr}
\begin{equation}
d_n
\sim
\frac{\alpha_s e}{4 \pi m_N}
\theta_{\rm SUSY}
,
\label{eq:nEDMCPVmass}
\end{equation}
where $\theta_{\rm SUSY}$ is the typical CP phase of the squark sector.
The loop function of the CP-odd mass is $O(1)$, so this contribution does not damp when the masses of supersymmetric particles increase.
By equating the current experimental data of neutron EDM $|d_n |< 1.8 \times 10^{-26} e$ cm \cite{Abel:2020gbr}, this yields a very strong constraint $|\theta_{\rm SUSY}| < 10^{-10}$.

According to our argument, however, the CP-odd quark mass (\ref{eq:CP-odd_mass}) becomes unphysical.
The leading contribution to the neutron EDM is then given by the quark EDM or chromo-EDM, and scales as
\begin{equation}
d_n
\sim
\frac{\alpha_s m_q e}{4 \pi M_{\rm SUSY}^2}
\theta_{\rm SUSY}
,
\label{eq:nEDMqEDM}
\end{equation}
where $M_{\rm SUSY}$ is the typical mass of supersymmetric particles.
By assuming $M_{\rm SUSY} \sim 1$ TeV, the constraint on the CP phase is relaxed to $|\theta_{\rm SUSY}| < 10^{-2}$.
This looser limit was also obtained from the axion mechanism which removes the effect of $\bar \theta$, but the crucial difference with our result is that we do not need to add a new (pseudo)scalar field in the theory.
Another minor difference is that the induced $\theta$-term generated by the linear axion term (\ref{eq:linearaxioncoupling}) does not exist anymore.
If we restrict to the CP violation of the quark sector, the phenomenological constraints are qualitatively similar between our result and that given by the axion mechanism.

The remarkable point of our result is that the $\theta$-term and the CP-odd masses of quarks, which are both dimension-4 operators do not contribute anymore to the observable CP violation.
These dimension-4 operators were so far problematic since they could avoid the decoupling of heavy new particle effects.
After our discussion, only operators with dimension-5 or higher in the flavor diagonal sector are relevant in probing the CP violation, so the decoupling of heavy mass scales, as we may see in Eq. (\ref{eq:nEDMqEDM}), is now correctly working.
We may in some sense say that the unobservability of the CP phases of the quark masses is also another sort of the resolution of the Strong CP problem.
At the same time, this means that the only source of CP violation in the SM is the CP phase of the Cabibbo-Kobayashi-Maskawa (CKM) matrix \cite{Kobayashi:1973fv}.

\section{Consistency with lattice gauge theory studies\label{sec:lattice}}

In this Section, we summarize the important lattice results related to the topology of nonabelian gauge theory, and inspect the consistency with our findings.

The strongest argument in favor of the observability of the topological charge of nonabelian gauge theory was so far the ``observation'' of instantons on lattice \cite{Chu:1994vi,Faccioli:2003qz,Athenodorou:2018jwu}.
We also note that the topological charge is equal to the number of chiral Dirac zero-modes due to Atiyah and Singer's index theorem (\ref{eq:atiyah-singer}), and consistency has been found between the topological charges calculated using both definitions on lattice \cite{Alexandrou:2017hqw}.
From these previous works, we admit that gauge configurations generated on lattice have instanton configurations, and that the topological charge is ``calculable''.
However, we point out that this result is just the amplitude, and not the amplitude squared which is observed by us in reality as probability.
If we take the correct sum of states with indefinite metric (see Sec. \ref{sec:BRST}), we would certainly find a null probability with the amplitude involving the topological charge, of course up to some systematic error due to the discretization and finite volume.
With this in mind, one might be tempted to project all gauge configurations onto one topological sector, to reduce the computational cost or for any other practical reasons, but more studies are required e.g. to certify the detailed balance of the Markov chain while generating the gauge ensemble.
Another important point is the good fit of gauge configurations generated on lattice with the instanton liquid model.
For this, we have to note that instantons that are only separated by finite distances (=correlated) are physical, which never results in an exact single instanton carrying topological charge, as discussed in Sec. \ref{sec:instanton}.

The nucleon EDM is the second topic to be investigated.
There were so far many lattice works trying to quantify the contribution of the $\theta$-term to the nucleon EDM.
The full simulation of QCD with the $\theta$-term is having the notorious sign problem, so the effect of $\theta$ is treated by perturbatively inserting either the gluonic topological charge density \cite{Shindler:2015aqa,Dragos:2019oxn,Alexandrou:2020mds} or the flavor singlet CP-odd quark mass operator (\ref{eq:CP-odd_mass}) \cite{Guo:2015tla}.
We could conclude that both are unphysical through our discussion (see Sec. \ref{sec:EDM}), so it is now meaningless in the context of particle physics phenomenology to calculate these quantities on lattice, but let us try to interpret the results of the previous works.
We point out that the topological charge density and the CP-odd mass of quarks are also the interpolating operators of the $\eta$' meson (and also $\eta$) \cite{Kuramashi:1994aj,Christ:2010dd,Fukaya:2015ara}.
In this work, we saw that the local topological charge density operator is physical.
This one, together with the CP-odd mass, should be considered as the insertion of the ``$\eta$'-pole'', which is CP-violating, thus contributing to the EDM.
Another point to be treated with care is that the EDM calculated on lattice is a path integral, a form factor, an amplitude (of scattering), and not the probability which is given by the amplitude squared summed over indefinite metric, as seen in the previous paragraph.
After correct summation, there should in principle be no observable effects.
It might also be possible that some contamination from the unphysical topological charge is present.
We note that the lattice calculation of the CP-odd quark mass contribution to the nucleon EDM may be used  to test the chiral perturbation theory, so it is not completely meaningless even if its result cannot be directly used in the analysis of physics beyond the SM.

Finally, let us also see the consistency of our work with lattice QCD studies of the $U(1)_A$ symmetry above the critical temperature $T_c$.
Recent lattice QCD calculations of meson correlators with good control of chiral symmetry are suggesting us that $U(1)_A$ is restored at $T > T_c$ \cite{Cossu:2013uua,Chiu:2013wwa,Brandt:2016daq,Ishikawa:2017nwl,Bazavov:2019www,Aoki:2020noz}.
These are in contrast to the usual statement that the axial $U(1)$ symmetry is explicitly broken by the chiral anomaly or instantons independently of the temperature.
On the other hand, the ``physical'' anomalous WTI (\ref{eq:physicalchiralWTI}) that we derived in Sec. \ref{sec:fermion} is having the unphysical gluonic surface term on the right-hand side, and the global violation due to the topology of gauge configurations is unphysical.
This means that the $U(1)_A$ symmetry suffers at most from the spontaneous chiral symmetry breaking, without explicit global symmetry breaking up to current quark masses.
The effective restoration of $U(1)_A$ above $T_c$ seen on lattice should then be true.
This downgrade of the $U(1)_A$ violation must certainly have an impact in the study of the QCD phase diagram or the cosmology of the early era which were based on the explicit violation, although we do not discuss them in this work.
It would also be interesting to inspect using lattice calculations the behavior of the $U(1)_A$ symmetry in color $SU(2)$ gauge theory with quarks at finite density, since we do not have there the sign problem.
A possible approach is to analyze the effect of the 't Hooft vertex by varying the chemical potential \cite{Suenaga:2022uqn}, but the existing lattice result is still incomplete due to the neglect of disconnected diagrams \cite{Murakami:2022lmq}.
We expect this $SU(2)$ analysis to be performed soon.

It was also recently claimed that the color fields of QCD with finite $\theta$-term is screened in the infrared region in lattice simulation \cite{Nakamura:2021meh}.
This result is apparently not inconsistent with ours at least at long distance in that we do not have color deconfinement, but a more detailed inspection is clearly needed to verify the consistency.

\section{Summary}

In this paper, we showed that the topological charge of nonabelian gauge theory is unphysical.
Physics that are affected by our conclusion are the followings:
\begin{itemize}

\item
Resolution of the Strong CP problem. Axions are not needed to resolve it.

\item
Topological $n$-point correlators do not probe the topology of gauge fields.
The vacuum energy does not depend on $\theta$.

\item
Chiral Dirac zero-modes become unphysical together with the topological charge according to Atiyah and Singer's index theorem: the $U(1)_A$ symmetry is a symmetry of QCD up to current quark masses.
$U(1)_A$ is thus spontaneously broken by the dynamics of QCD, and $\eta'$ is the ninth pseudo-NG boson.

\item
Isolated instantons are unphysical. 
The correlated ones, separated by finite distances, are observable, but do not change the topology of gauge fields.

\item
Baryogenesis via sphaleron is forbidden.

\item
Chiral magnetic effect due to the topological chiral chemical potential is forbidden.

\item
CP phases of quark masses are rotated away after quark field redefinition, and they become unphysical.
Dimension-4 CP violation is unobservable except the CP phase of the CKM matrix which is the only source of CP nonconservation in the SM.

\end{itemize}

We also list physics that are not affected by our conclusion:
\begin{itemize}
\item
Axion dark matter scenario is not constrained and phenomenological analyses of axions, as well as axion-like particles (ALPs) are still required. 
The axion-gauge interaction (\ref{eq:axionlagrangian}) is in general allowed, so the studies of axions and ALPs that were performed so far should rather be considered as general properties to be phenomenologically checked when we introduce a new scalar boson, just like the Higgs potential or the hierarchical problem associated with the cutoff.
We have to keep in mind that the axion does not contribute anymore to the resolution of the Strong CP problem, so theories with light scalars with the coupling (\ref{eq:axionlagrangian}) will become specialized to explain scenarios with ultra-light dark matter (and perhaps also to the inflation, if the axion effective potential is very ``slow-roll'').

\item
Since the $\theta$-term is unphysical, parity and CP symmetries seem to be protected in nonabelian gauge theory and to have a special meaning.

\item
Instanton liquid models can be physical, but the topological contribution to fermions (e.g. KMT interaction) has to be removed.

\item
Lattice calculations will probably be simplified, since taking into account topological sectors affects the computational cost in many cases.
However, more detailed inspections are needed to establish whether the topology is really unphysical in discretized formulations, and the technical validity, especially in the Montecarlo sampling.

\item
Some remark: the Skyrme model, in which baryons are generated by the topology of NG boson fields, is not affected by our conclusion since the background field is not a gauge field.

\end{itemize}

\section*{Acknowledgement}

The author thanks Yoshikazu Hagiwara, David Dudal, and Hiroaki Abuki for useful discussions.
This work was supported by Daiko Foundation.

\appendix

\section{BRST transform of topological current\label{sec:appendixBRST}}

The BRST transform of Eq. (\ref{eq:topologicalcurrent}) is
\begin{eqnarray}
\delta_B K_\mu
&=&
\delta_B 
\left[
\frac{\alpha_s}{4\pi}
\epsilon_{\mu \nu \rho \sigma}
\Biggl\{
A^\nu_a \partial ^\rho A_a^{\sigma }
+\frac{1}{3}
g_s
f_{abc}
A^\nu_a A^\rho_b A_c^\sigma 
\Biggr\}
\right]
\nonumber\\
&=&
\frac{\alpha_s}{4\pi}
\epsilon_{\mu \nu \rho \sigma}
\Biggl[
(\partial^\nu c_a) \partial ^\rho A_a^{\sigma }
+g_s f_{abc} A^\nu_b c_c \partial ^\rho A_a^{\sigma }
+A^\nu_a \partial ^\rho (\partial^{\sigma } c_a)
+g_s f_{abc} A^\nu_a \partial ^\rho (A_b^{\sigma } c_c)
\nonumber\\
&& \hspace{4em}
+\frac{1}{3}
g_s
f_{abc}
(\partial^\nu c_a) A^\rho_b A_c^\sigma 
+\frac{1}{3}
g_s
f_{abc}
A^\nu_a (\partial^\rho c_b) A_c^\sigma 
+\frac{1}{3}
g_s
f_{abc}
A^\nu_a A^\rho_b (\partial^\sigma c_c ) 
\nonumber\\
&& \hspace{4em}
+\frac{1}{3}
g_s^2
f_{abc}
f_{ade}
A^\nu_d c_e A^\rho_b A_c^\sigma 
+\frac{1}{3}
g_s^2
f_{abc}
f_{bde}
A^\nu_a A^\rho_d c_e A_c^\sigma 
+\frac{1}{3}
g_s^2
f_{abc}
f_{cde}
A^\nu_a A^\rho_b A_d^\sigma c_e
\Biggr]
\nonumber\\
&=&
\frac{\alpha_s}{4\pi}
\epsilon_{\mu \nu \rho \sigma}
\Bigl[
(\partial^\nu c_a) \partial ^\rho A_a^{\sigma }
+g_s f_{abc} A^\nu_b c_c \partial ^\rho A_a^{\sigma }
+g_s f_{abc} A^\nu_a (\partial^\rho A_b^{\sigma }) c_c
+g_s f_{abc} A^\nu_a A_b^{\sigma } (\partial^\rho c_c)
+
g_s
f_{abc}
(\partial^\nu c_a) A^\rho_b A_c^\sigma 
\Bigr]
\nonumber\\
&=&
\frac{\alpha_s}{4\pi}
\epsilon_{\mu \nu \rho \sigma}
(\partial^\nu c_a) \partial ^\rho A_a^{\sigma }
.
\end{eqnarray}
In the third equality, we used the Jacobi identity
\begin{equation}
f_{abc}f_{bde}+f_{ebc}f_{dba}+f_{dbc}f_{abe}=0
,
\label{eq:jacobiidentity}
\end{equation}
to erase the $O(A^3c)$ terms.

The BRST transform of the topological anti-ghost current (\ref{eq:topologicalantighost}):
\begin{eqnarray}
\delta_B \bar {\cal C}_\mu
\ =\ {\cal B}_\mu
&=&
\delta_B
\Biggl[
\frac{\alpha_s}{8\pi}
\epsilon_{\mu \nu \rho \sigma}
(\partial^{-2} \partial^\nu \bar c_a) 
g_s f_{abc} A_b^\rho A_c^\sigma
\Biggr]
\nonumber\\
&=&
\frac{\alpha_s}{8\pi}
\epsilon_{\mu \nu \rho \sigma}
g_s f_{abc} 
\Bigl[
(\partial^{-2} \partial^\nu B_a) 
A_b^\rho A_c^\sigma
+
(\partial^{-2} \partial^\nu \bar c_a) 
(\partial^\rho c_b )
A_c^\sigma
+
(\partial^{-2} \partial^\nu \bar c_a) 
g_s f_{bde}
A_d^\rho c_e
A_c^\sigma
\nonumber\\
&& \hspace{12.5em}
+
(\partial^{-2} \partial^\nu \bar c_a) 
A^\rho_b
(\partial^\sigma c_c )
+
(\partial^{-2} \partial^\nu \bar c_a) 
A_b^\rho 
g_s f_{cde}
A_d^\sigma
c_e
\Bigr]
\nonumber\\
&=&
\frac{\alpha_s}{8\pi}
\epsilon_{\mu \nu \rho \sigma}
g_s f_{abc} 
\Bigl[
(\partial^{-2} \partial^\nu B_a) 
A_b^\rho A_c^\sigma
+
2(\partial^{-2} \partial^\nu \bar c_a) 
(\partial^\rho c_b )
A_c^\sigma
+
2 g_s 
(\partial^{-2} \partial^\nu \bar c_a) 
f_{bde}
A_d^\rho c_e
A_c^\sigma
\Bigr]
\nonumber\\
&=&
\frac{\alpha_s}{8\pi}
\epsilon_{\mu \nu \rho \sigma}
g_s f_{abc} 
\Bigl[
(\partial^{-2} \partial^\nu B_a) 
A_b^\rho A_c^\sigma
+
2(\partial^{-2} \partial^\nu \bar c_a) 
(\partial^\rho c_b )
A_c^\sigma
+
g_s 
(\partial^{-2} \partial^\nu \bar c_a) 
f_{bde}
A_d^\rho 
A_e^\sigma
c_c
\Bigr]
.
\end{eqnarray}
In the fifth identity, we again used the Jacobi identity (\ref{eq:jacobiidentity}).

\end{document}